\documentclass[english]{extarticle}
\usepackage[T1]{fontenc}
\usepackage[latin9]{inputenc}
\usepackage{geometry}
\usepackage{color}
\usepackage{babel}
\usepackage{units}
\usepackage{amsmath}
\usepackage{amsthm}
\usepackage{amssymb}
\usepackage{setspace}
\usepackage[unicode=true,pdfusetitle,
 bookmarks=true,bookmarksnumbered=false,bookmarksopen=false,
 breaklinks=true,pdfborder={0 0 1},backref=false,colorlinks=false]
 {hyperref}
\usepackage{breakurl}

\makeatletter

\providecommand{\tabularnewline}{\\}

\numberwithin{equation}{section}
\numberwithin{figure}{section}
\newcommand{\lyxaddress}[1]{
\par {\raggedright #1
\vspace{1.4em}
\noindent\par}
}

\@ifundefined{date}{}{\date{}}
\makeatother

\begin{document}

\title{\textbf{A new development in quantum field equations of dyons}}

\author{\textbf{B. C. Chanyal }}
\maketitle

\lyxaddress{\begin{center}
\textit{Department of Physics,}\\
\textit{G. B. Pant University of Agriculture \& Technology,}\\
\textit{ Pantnagar-263145 (Uttarakhand), India }\\
\textit{E-mail: bcchanyal@gmail.com, bcchanyal@gbpuat.ac.in}
\par\end{center}}
\begin{abstract}
In this study, we describe a novel approach to quantum phenomena of
the generalized electromagnetic fields of dyons with quaternionic
analysis. Starting with quaternionic quantum wave equations, we have
established a quantized condition for time coordinate that transforms
microscopic to macroscopic fields. In view of classical electromagnetic
field equation, we propose a new set of quantized Proca-Maxwell's
equations for dyons. Furthermore, a quantized form of four-currents
densities and the quantized Lorentz gauge conditions, respectively
for electric and magnetic potentials of dyons are obtained. We have
established the new quantized continuity equations for electric and
magnetic densities of dyons which associated with a torque density
result from the two spin states. The quantized Klein-Gordon like field
equations and the unified quaternionic electromagnetic potential wave
equations for massive dyons are demonstrated. Moreover, we investigate
the quaternionic quantized relativistic Dirac field equations for
massive dyons, which indicated that there will be the existence of
antiparticle of dyons called \textit{antidyons}.

\textbf{Keywords:} quaternion, dyons and anti-dyons, Maxwell equations,
quantized Lorentz gauge, Dirac field equation,

\textbf{PACS:} 02.10.De, 03.50.De, 03.65.-w
\end{abstract}

\section{Introduction}

The generalized Maxwell's equations describe a complete and valuable
description to the classical behavior of electric and magnetic fields.
At the end of the twentieth-century, the asymmetry between electric
and magnetic field equations became very clear with the formulation
of Maxwell's equations. In 1931, Dirac \cite{key-1,key-2} investigated
a new particle called 'magnetic monopoles' and also generalized the
usual electrodynamics. The basic idea for magnetic monopoles is that,
it will be stable particles, which carry magnetic charges like electric
charges for electrons, ought to exist has proved to be remarkably
durable. The existence of particles with the magnetic charge (or monopole)
\cite{key-1,key-2} implies that the electric charge must be integer
multiples of a fundamental units. Beside, a quantization of electric
charges have been actually observed in nature. Dirac also have confirmed
the constancy of quantum physics with unique Dirac equation, particularly
in solving the difficulties associated with the string singularity
identified by Dirac \cite{key-2,key-3}. Thus, we can say that the
magnetic monopole would symmeterize in Maxwell's equations, but there
would be some numerical asymmetry. These types of discrepancy could
be the basis for the introduction of what we may call the 'classical
magnetic monopole'. In this classical formulation there was no prediction
for the magnetic-monopole mass. Rather, a kind of rule was established,
and also assumed that the radius of classical electron may be equal
to the radius of classical monopole from which one has $m_{0}^{g}=\nicefrac{\mathtt{g}_{d}^{2}m_{0}^{e}}{e^{2}}\approx4700m_{0}^{e}\approx2.4GeV\,(\text{approx}).$
This ingenious suggestions with the existence of magnetic-monopole
gave rise to considerable literature \cite{key-4}-\cite{key-10}
that subjected to predict the mass, size, spin, parity and other quantum
properties of monopoles. In the case of the existence of magnetic
monopole, Schwinger \cite{key-11}-\cite{key-13} formulated a novel
relativistic quantum field theory of magnetic charges which maintained
the complete symmetry between electric and magnetic fields. The Dirac's
quantization condition governed the product of electric and magnetic
charge to integer values. The combined theories of Schwinger \cite{key-11}-\cite{key-13}
and Zwanziger \cite{key-14} identify the theory of dyons (i.e., particles
carrying simultaneous existence of electric and magnetic charges).
The quantized theory, though explains to some extent the negative
experimental results in search of magnetic monopole, and required
to maintained the rotational symmetry which was violated due to existence
of singular lines which comes from the solution of vector potential
around a monopole. Peres \cite{key-15} discussed the controversial
nature \cite{key-16} of these singular lines of monopole \cite{key-1,key-17}
and then the charged quantization condition has taken place in purely
group theoretical manner. For electromagnetic fields, Maxwell equations
are considered to be the classical description and the quantization
of these classical fields gives a relativistic covariant quantum description.
Furthermore, Bialynicki-Birula \cite{key-18,key-19} and Sipe \cite{key-20}
independently explained that Maxwell equations are the quantum description
of electromagnetic fields at the first quantum level and the Fermat
principle for light forms the classical description of electromagnetic
fields similar to the Hamilton\textquoteright s variational principle
for particles with mass. 

In quaternionic space-time, Rajput et. al. \cite{key-21} proposed
a unified theory of generalized electromagnetic and gravitational
fields associated with massive dyons. Many other authors \cite{key-22,key-23}
discussed the development of classical and quantum theory of electrodynamics
in case of experimental data. Chanyal et al. \cite{key-24}-\cite{key-29}
have studied classical field equations and conservation laws of dyons
in terms of octonion algebra and explained the corresponding generalized
Dirac Maxwell\textquoteright s equations with the equation of motion
in compact manner. Moreover, Arbab \cite{key-30,key-31} derived the
role of quaternionic quantum mechanics in case of quantized Maxwell
equations. Besides, in literature \cite{key-32}-\cite{key-41}, the
reformulation of the field equations of magnetic monopoles and massive
dyons have been discussed in terms of hyper complex numbers including
quaternions, octonions and sedenions. Recently, Chanyal \cite{key-42}
independently proposed a covariant theory of relativistic quantum
mechanics for dyons wave propagation in terms of quaternionic formalism.
Keeping in mind the recent work in quaternionic quantum mechanics,
we establish a novel approach to quantum phenomena in the generalized
electromagnetic fields of dyons with quaternionic analysis. Starting
with quaternionic four dimensional structure and its properties, we
propose the relativistic quantized wave equations. Since classical
theory is a approximation of quantum theory, in this study, we establish
a quantized condition for time coordinate that transforms quantum
to classical field. We describe a compact set of quantized Proca-Maxwell's
equations for dyons. Furthermore, a quantized form of four potentials,
four-currents densities and also the quantized Lorentz gauge conditions,
respectively for electric and magnetic potentials of dyons are obtained.
We obtain a new quantized continuity equation for generalized electric
and magnetic four-densities of dyons which associated with a torque
density result from the two spin states. The quantized Klein-Gordon
like field equations and the unified quaternionic electromagnetic
potential wave equations for massive dyons are also obtained. Moreover,
we investigate the quaternionic quantized relativistic Dirac field
equations for massive dyons, which indicated that there will be the
existence of antiparticle of dyons called antidyons. Our quaternionic
quantum theory will help to establish the existence for various massive
particles.

\section{The quaternion}

In mathematics, a geometric algebra is the largest possible associative
division algebra that integrates all algebraic systems (viz., algebra
of complex numbers, vector algebra, matrix algebra, quaternion algebra,
etc.) into a coherent mathematical language. Interestingly, after
the real and the complex algebras, quaternion \cite{key-43} is first
hyper-complex division algebra. It has more significant impacts on
mathematics and theoretical physics. The quaternion algebra (called
\textit{q-algebra}), defined by $\mathbb{Q}$, is a four-dimensional
algebra \cite{key-43,key-44} over the field of real numbers. A real
quaternion $\mathbb{Q}$ and its quaternionic conjugate $\overline{\mathbb{Q}}$
may be expressed as
\begin{align}
\mathbb{Q}\,= & \left(q_{0},\,\vec{q}\right)\equiv q_{0}e_{0}+\sum_{j=1}^{3}q_{j}e_{j},\,\,\,(\forall\,j=1,2,3)\label{eq:1}\\
\overline{\mathbb{Q}}\,= & \left(q_{0},\,-\vec{q}\right)\equiv q_{0}\bar{e}_{0}+\sum_{j=1}^{3}q_{j}\bar{e}_{j}=q_{0}e_{0}-\sum_{j=1}^{3}q_{j}e_{j}\,,\label{eq:2}
\end{align}
where $\left(q_{0}e_{0}\right)$ and $\left(q_{j}e_{j}\right)\,\forall j=1,2,3$
are the scalar and vector parts of q-algebra. The q-units defined
by $(e_{0},e_{1},e_{2},e_{3})$ are known as the quaternion basis.
The properties of q-algebra are given by: 
\begin{alignat}{1}
e_{0}^{2} & =\,e_{0}=\,1\,\,,\,\,e_{j}^{2}=-e_{0},\,\,\bar{e}_{j}=-e_{j},\,\bar{e}_{0}=e_{0}\nonumber \\
e_{0}e_{j} & =\,e_{j}e_{0}=\,e_{j}\,\,,\,\,e_{i}e_{j}=-\delta_{ij}+\varepsilon_{ijk}e_{k}\,\,\,(\forall\,i,j,k=1,2,3)\,\,.\label{eq:3}
\end{alignat}
In equation (\ref{eq:3}) we have used $\delta_{ij}$ as delta symbol,
$\bar{e}_{j}$ as quaternionic conjugate basis and $\varepsilon_{ijk}$
as the Levi Civita three index symbol. Since, the q-algebra is an
associative as well as non-commutative algebra in nature and holds
the usual distribution law, i.e., $(e_{j}e_{k})e_{l}=e_{j}(e_{k}e_{l})\,\forall\,i,j,k=1,2,3$.
Moreover, we may distinguish the real part of the quaternion $q_{0}$
as,
\begin{alignat}{1}
\mathfrak{Re}\,\mathbb{Q} & =\,\,\frac{1}{2}(\overline{\mathbb{Q}}+\mathbb{Q})=\,\,q_{0}\,\,.\label{eq:4}
\end{alignat}
If $\mathfrak{Re}\,\mathbb{Q}=0$ or $q_{0}=0$, then we may found
the imaginary part of the q-algebra and the condition become $\mathbb{Q}=-\overline{\mathbb{Q}}$,
where
\begin{alignat}{1}
\mathfrak{Im}\,\mathbb{Q}=\,\frac{1}{2}(\mathbb{Q}-\overline{\mathbb{Q}}) & =\,\left(q_{1}e_{1}+q_{2}e_{2}+q_{3}e_{3}\right)\,\,.\label{eq:5}
\end{alignat}
Furthermore, the summation and the multiplication of any two quaternionic
variables are given by
\begin{align}
\mathbb{P}\pm\mathbb{Q}= & \left(p_{0}\pm q_{0}\right)+\left(\vec{p}\pm\vec{q}\right)\nonumber \\
= & \left(p_{0}\pm q_{0}\right)e_{0}+\left(p_{1}\pm q_{1}\right)e_{1}+\left(p_{1}\pm q_{1}\right)e_{2}+\left(p_{1}\pm q_{1}\right)e_{3}\,\,,\label{eq:6}
\end{align}
\begin{align}
\mathbb{P}\,\mathbb{Q}= & \left[p_{0}+\vec{p}\right]\left[q_{0}+\vec{q}\right]=p_{0}q_{0}+p_{0}\vec{q}+q_{0}\vec{p}-\left(\vec{p}\cdot\vec{q}\right)+\left(\vec{p}\times\vec{q}\right)\,\,,\label{eq:7}
\end{align}
where ($\cdot$) and ($\times$) are indicated dot and cross products
like the usual three-dimensional scalar and vector products. In equation
(\ref{eq:7}), we should notice that the q-multiplication $\mathbb{P}\,\mathbb{Q}\neq\mathbb{Q}\,\mathbb{P}$,
because $\vec{p}\times\vec{q}\neq-\vec{q}\times\vec{p}$. Rather,
the norm of a q-algebra become $N(\mathbb{Q})=\mathbb{Q}\,\overline{\mathbb{Q}}=\overline{\mathbb{Q}}\,\mathbb{Q}$
$=\sum_{\alpha=0}^{3}\,\mathbb{Q}_{\alpha}^{2}$, and the inverse
of a q-algebra (non-zero norm) become $\mathcal{\mathbb{Q}}^{-1}=\frac{\overline{\mathbb{Q}}}{N(\mathcal{\mathbb{Q}})}\,;\,\,\mathcal{\mathbb{Q}}\,\mathcal{\mathbb{Q}}^{-1}$
$=\mathcal{\mathbb{Q}}^{-1}\,\mathcal{\mathbb{Q}}=1.$ The norm of
a q-algebra is zero if $\mathcal{\mathbb{Q}}=0$, and is always positive
otherwise. Thus keeping in mind the q-properties, in the next sections
we shall implement q-algebra in relativistic quantum mechanics.

\section{Quaternion approach to relativistic quantum field equations}

In physics point of view, the four independent quantities such as
the four-position, four-momentum, four-force, four-potential and four-current
etc. are suitable to express by q-algebra. In other words, the quaternionic
approach is a such type of novel approach where we generalize and
explain beautifully to the quantum behavior of relativistic equations.
In order to discuss the quaternionic quantum mechanics (qQM), we may
start with the momentum eigen value equation \cite{key-30,key-31},
\begin{align}
\breve{P}\,\breve{\varPsi}\,\,= & \,\,\varUpsilon\,\breve{\varPsi}\,,\,\,\,\,\,\left(\breve{P},\,\breve{\varPsi}\in\mathbb{Q}\right)\,,\label{eq:8}
\end{align}
where $\varUpsilon\sim\left(m_{0}c\right)$ is the eigen value of
quaternionic four-momentum operator $\breve{P}=\left\{ \overrightarrow{p},\,\frac{i}{c}E\right\} $,
the eigen function $\breve{\varPsi}=\left\{ \overrightarrow{\varPsi},\,\frac{i}{c}\varPsi_{0}\right\} $
is quaternionic four-wave function, $m_{0}$ is the mass of particle
(dyon) and $c$ is the speed of light. By using the quaternionic product
for two variables given by (\ref{eq:7}) the quaternionic momentum
eigen value equation (\ref{eq:8}) leads to
\begin{align}
\underset{\text{vector\,part}}{\underbrace{\left(\frac{i}{c}E\right)\overrightarrow{\varPsi}+\overrightarrow{p}\left(\frac{i}{c}\varPsi_{0}\right)+\overrightarrow{p}\times\overrightarrow{\varPsi}}}\,\,,\,\,\,\,\,\underset{\text{scalar part }}{\underbrace{-\frac{E}{c^{2}}\varPsi_{0}-\overrightarrow{p}\cdot\overrightarrow{\varPsi}}}= & \,\,m_{0}c\,\left\{ \overrightarrow{\varPsi},\,\,\,\frac{i}{c}\varPsi_{0}\right\} \,.\label{eq:9}
\end{align}
Now, the quantum equations for a moving particle can be expressed
by \cite{key-45,key-46},
\begin{align}
\overrightarrow{\nabla}\cdot\overrightarrow{\varPsi}-\frac{1}{c^{2}}\frac{\partial\varPsi_{0}}{\partial t}-\frac{m_{0}}{\hbar}\varPsi_{0}= & \,\,0\,,\label{eq:10}\\
\overrightarrow{\nabla}\varPsi_{0}-\frac{\partial\overrightarrow{\varPsi}}{\partial t}-\frac{m_{0}c^{2}}{\hbar}\overrightarrow{\varPsi}= & \,\,0\,,\label{eq:11}\\
\overrightarrow{\nabla}\times\overrightarrow{\varPsi}= & \,\,0\,,\label{eq:12}
\end{align}
where the momentum and energy operator are $\overrightarrow{p}=-i\hbar\overrightarrow{\nabla}$
and $E=i\hbar\frac{\partial}{\partial t}$. The beauty of the equation
(\ref{eq:8}) is that, it exhibited the well-known Dirac equation
for massive dyons, i.e., $\left(i\hbar\gamma^{\nu}\partial_{\nu}-m_{0}c\right)\breve{\varPsi}=0$
if we put the four dimensional quaternionic momentum operator $\breve{P}\longrightarrow\left(i\hbar\gamma^{\nu}\partial_{\nu}\right)$.
We may established the second order quantum differential equations
by operating $\overrightarrow{\nabla}$ to the both side from left
on equations (\ref{eq:10}-\ref{eq:12}), i.e.
\begin{align}
\nabla^{2}\overrightarrow{\varPsi}-\frac{1}{c^{2}}\frac{\partial^{2}\overrightarrow{\varPsi}}{\partial t^{2}}-2\left(\frac{m_{0}}{\hbar}\right)\frac{\partial\overrightarrow{\varPsi}}{\partial t}-\frac{m_{0}^{2}c^{2}}{\hbar^{2}}\overrightarrow{\varPsi}= & \,\,0\,,\label{eq:13}\\
\nabla^{2}\varPsi_{0}-\frac{1}{c^{2}}\frac{\partial^{2}\varPsi_{0}}{\partial t^{2}}-2\left(\frac{m_{0}}{\hbar}\right)\frac{\partial\varPsi_{0}}{\partial t}-\frac{m_{0}^{2}c^{2}}{\hbar^{2}}\varPsi_{0}= & \,\,0\,,\label{eq:14}\\
\overrightarrow{\nabla}\times\left(\overrightarrow{\nabla}\times\overrightarrow{\varPsi}\right)\equiv\overrightarrow{\nabla}\left(\overrightarrow{\nabla}\cdot\overrightarrow{\varPsi}\right)-\nabla^{2}\overrightarrow{\varPsi}= & \,\,\,0\,.\label{eq:15}
\end{align}
Equation (\ref{eq:13}) represents the generalized quaternionic quantum
wave equation for massive vector field ($\overrightarrow{\varPsi}$)
which may visualized by Dirac field, whereas equation (\ref{eq:14})
represents to massive scalar field (like as the Klein-Gordon field
$\varPsi_{0}$). Equation (\ref{eq:15}) shows the quaternionic condition
for the triple vector multiplication. The unified structure of quaternionic
quantum wave equation is expressed by 
\begin{align}
\nabla^{2}\breve{\varPsi}-\frac{1}{c^{2}}\frac{\partial^{2}\breve{\varPsi}}{\partial t^{2}}-2\left(\frac{m_{0}}{\hbar}\right)\frac{\partial\breve{\varPsi}}{\partial t}-\frac{m_{0}^{2}c^{2}}{\hbar^{2}}\breve{\varPsi}= & \,\,0\,\,.\label{eq:16}
\end{align}
Here we should notice that, in quaternionic quantum wave equation
the additional term $\left(\frac{2m_{0}}{\hbar}\right)\frac{\partial\breve{\varPsi}}{\partial t}$
describes the time dependent source called the damping terms resulting
from the inertia of the massive particle. A beauty of quaternionic
unified wave equation is that we may transform this quantum wave equation
into classical wave equation if we eliminate the damping term by introducing
a new time coordinate $\tau$ as
\begin{align}
\frac{\partial}{\partial\tau}\,\longmapsto\, & \left(\frac{\partial}{\partial t}+\frac{1}{\triangle\mathfrak{t}}\right)\,,\,\,\,\,\,\text{where \,\,}\triangle\mathfrak{t}=\frac{\hbar}{m_{0}c^{2}}\approx\frac{\hbar}{\triangle E_{k}}\,.\label{eq:17}
\end{align}
This is the necessary condition for to transform macroscopic to microscopic
field in quaternionic space-time. Then the ordinary macroscopic wave
equations becomes,
\begin{align}
\nabla^{2}\overrightarrow{\varPsi}-\frac{1}{c^{2}}\frac{\partial^{2}\overrightarrow{\varPsi}}{\partial\tau^{2}}= & \,\,0\,,\,\,\,\,\:\Longrightarrow\,\,\lozenge\overrightarrow{\varPsi}=\,0\,,\label{eq:18}\\
\nabla^{2}\varPsi_{0}-\frac{1}{c^{2}}\frac{\partial^{2}\varPsi_{0}}{\partial\tau^{2}}= & \,\,0\,,\,\,\,\,\:\Longrightarrow\,\,\lozenge\,\varPsi_{0}=\,0\,,\label{eq:19}
\end{align}
where $\lozenge=\left(\nabla^{2}-\frac{1}{c^{2}}\frac{\partial^{2}}{\partial\tau^{2}}\right)$
denotes classical D\textquoteright{} Alembert operator for time coordinate
$\tau$. \\
Now, in view of the physical significant of generalized quantum wave
equation, we may consider a general plane wave solution of equation
(\ref{eq:16}) as
\begin{align}
\breve{\varPsi}=\,\,\xi\,\exp\,i(\omega t-\overrightarrow{k}\cdot\overrightarrow{r}),\,\,\,\, & \text{where\,\,}\xi=\text{constant}.\label{eq:20}
\end{align}
By substituting (\ref{eq:20}) in equation (\ref{eq:16}), we may
found two Dirac like solutions respectively for positive energy with
particle energy $\hbar\omega^{+}$ and negative energy with antiparticle
energy $\hbar\omega^{-}$ \cite{key-42}. Furthermore, the group and
phase velocities respectively, $v_{g}$ and $v_{p}$ can be expressed
by the relations $v_{g}=\,\,\frac{\partial\omega^{\pm}}{\partial k}=\pm\,c\,,$
and $v_{p}=\,\,\frac{\omega^{\pm}}{\overrightarrow{k}}=\frac{im_{0}c^{2}}{\hbar\overrightarrow{k}}\pm\,c\,.$

\section{Quaternionic classical electromagnetic field of dyons}

In order to write the classical electromagnetic field of dyons \cite{key-47},
let us start with the quaternionic four-vector representation with
an imaginary fourth component of Euclidean structure $(+,+,+,-)$.
The quaternionic four dimensional world vector $\mathbb{X}$ and the
four gradiant $\mathbb{D}$ may then be defined as
\begin{align}
\mathbb{X}\left(e_{1},\,e_{2},\,e_{3},\,\,\,e_{0}\right):= & \,\left\{ x,\,y,\,z,\,-ict\right\} \,,\label{eq:21}\\
\mathbb{D}\left(e_{1},\,e_{2},\,e_{3},\,\,\,e_{0}\right):= & \,\left\{ \frac{\partial}{\partial x},\,\frac{\partial}{\partial y},\,\frac{\partial}{\partial z},\,-\frac{\partial}{i\partial(ct)}\right\} \,\,.\label{eq:22}
\end{align}
As such, we may write a quaternionic four-vector $\mathbb{X}^{\nu}$
and its quaternion conjugate $\bar{\mathbb{X}^{\nu}}$ in terms of
following covariant form,
\begin{align}
\mathbb{X}^{\nu}= & \,\,(x^{0}e_{0}+x^{1}e_{1}+x^{2}e_{2}+x^{3}e_{3})\,,\label{eq:23}\\
\bar{\mathbb{X}^{\nu}}= & \,\,(x^{0}e_{0}-x^{1}e_{1}-x^{2}e_{2}-x^{3}e_{3})\,,\,\,\,\,\,\,\,\,\,\,\,\,\,\,(\mathbb{X}^{\nu},\,\bar{\mathbb{X}^{\nu}}\in\,\mathbb{Q})\,,\label{eq:24.}
\end{align}
and the quaternionic equation of spherical surface is governed as
\begin{align}
\sum_{\nu=0}^{3}\mathbb{X}^{\nu}\bar{\mathbb{X}}_{\nu}\,\equiv\, & (x^{0}x_{0}+x^{1}x_{1}+x^{2}x_{2}+x^{3}x_{3})\,\simeq-c^{2}t^{2}+x^{2}+y^{2}+z^{2}\,.\label{eq:25}
\end{align}
Similarly, the classical D\textquoteright{} Alembert operator defined
by$\square$, can be expressed as
\begin{align}
\square=\,\,\mathbb{D}\bar{\mathbb{D}}\,\,=\bar{\mathbb{D}}\mathbb{D}\,\,\equiv & \,\,\frac{\partial^{2}}{\partial x^{2}}+\frac{\partial^{2}}{\partial y^{2}}+\frac{\partial^{2}}{\partial z^{2}}-\frac{1}{c^{2}}\frac{\partial^{2}}{\partial t^{2}}\nonumber \\
= & \,\,\left(\nabla^{2}-\frac{1}{c^{2}}\frac{\partial^{2}}{\partial t^{2}}\right)\,,\label{eq:26}
\end{align}
where $\bar{\mathbb{D}}$ is the quaternionic conjugate of the differential
operator $\mathbb{D}$. Since, q-algebra has four dimensional structure,
thus in a four-dimensional theory, a \textit{dyon} is defined to a
particle in which both electric and magnetic charges are taken place.
In the high dimensional theories viz. grand unified theory (GUT) and
super-string theory (SST) predicted to the existence of both magnetic
monopoles and dyons. Thus, a dyon constitutes two four-potentials
defined by following quaternionic form \cite{key-47,key-48}:
\begin{align}
\mathbb{A}\left(e_{1},\,e_{2},\,e_{3},\,\,\,e_{0}\right)= & \,\left\{ A_{x},\,A_{y},\,A_{z},-\frac{i}{c}\phi_{e}\right\} \,,\label{eq:27}\\
\mathbb{B}\left(e_{1},\,e_{2},\,e_{3},\,\,\,e_{0}\right)= & \,\left\{ B_{x},\,B_{y},\,B_{z},-\frac{i}{c}\phi_{m}\right\} \,,\label{eq:28}
\end{align}
where $\mathbb{A}\,\text{and }\mathbb{B}$ are electric and magnetic
four-potentials and their real and imaginary parts constituent a generalized
potential of dyons, i.e.
\begin{align}
\mathbb{V}^{\nu}\left\{ A^{\nu},\,B^{\nu}\right\}  & =\,\,\left(A^{\nu}+i\,B^{\nu}\right)\,\,,\,\,\,(\nu=0,1,2,3)\,.\label{eq:29}
\end{align}
As such, the quaternionic form of two four-currents of dyons, respectively,
\begin{align}
\mathbb{J}\left(e_{1},\,e_{2},\,e_{3},\,\,\,e_{0}\right)= & \,\left\{ J_{x},\,J_{y},\,J_{z},-ic\rho_{e}\right\} \,,\label{eq:30}\\
\mathbb{K}\left(e_{1},\,e_{2},\,e_{3},\,\,\,e_{0}\right)= & \,\left\{ K_{x},\,K_{y},\,K_{z},-ic\rho_{m}\right\} \,,\label{eq:31}
\end{align}
constitute a generalized current-source of dyons. Correspondingly,
the quaternionic generalized current defined by
\begin{align}
\mathbb{J}^{\nu}\left\{ J^{\nu},\,K^{\nu}\right\}  & =\,\,\left(J^{\nu}+i\,K^{\nu}\right)\,.\label{eq:32}
\end{align}
Similarly, the quaternionic representation of energy-momentum four-vector
of moving dyons become
\begin{align}
\mathbb{P}\left(e_{1},\,e_{2},\,e_{3},\,\,\,e_{0}\right)= & \,\left\{ p_{x},\,p_{y},\,p_{z},-\frac{i}{c}E\right\} \,.\label{eq:33}
\end{align}
Now, we may write the electromagnetic classical field equations called
generalized Dirac Maxwell's (GDM) equations for dyons in terms of
following symmetric nature,
\begin{align}
\overrightarrow{\nabla}\cdot\overrightarrow{\mathcal{E}} & +\frac{\partial\Lambda_{e}}{\partial t}-\frac{\rho_{e}}{\varepsilon_{0}}=0\,,\nonumber \\
\overrightarrow{\nabla}\cdot\overrightarrow{\mathcal{H}} & +\frac{\partial\Lambda_{m}}{\partial t}-\mu_{0}\rho_{m}=0\,,\nonumber \\
\overrightarrow{\nabla}\times\overrightarrow{\mathcal{E}} & +\frac{\partial\overrightarrow{\mathcal{H}}}{\partial t}+\overrightarrow{\nabla}\Lambda_{m}+\mu_{0}\overrightarrow{K}=0\,,\nonumber \\
\overrightarrow{\nabla}\times\overrightarrow{\mathcal{H}} & -\frac{1}{c^{2}}\frac{\partial\overrightarrow{\mathcal{E}}}{\partial t}+\overrightarrow{\nabla}\Lambda_{e}-\mu_{0}\overrightarrow{J}=0\,.\label{eq:34}
\end{align}
where $\overrightarrow{\mathcal{E}}\longmapsto\left(e_{1}\mathcal{E}_{x},\,e_{2}\mathcal{E}_{y},\,e_{3}\mathcal{E}_{z}\right)$
and $\overrightarrow{\mathcal{H}}\longmapsto\left(e_{1}\mathcal{H}_{x},\,e_{2}\mathcal{H}_{y},\,e_{3}\mathcal{H}_{z}\right)$
are denoted the quaternionic electric and the magnetic field vectors;
$\rho_{e}$ and $\rho_{m}$ are the charge source density due to electric
and magnetic charge (monopole); $\varepsilon_{0}$ is the free space
permittivity and $\mu_{0}$ is the permeability to the case of free
space. Here $\Lambda_{e}$ and $\Lambda_{m}$ are defined the relativistic
Lorentz gauge, respectively for electric and magnetic charges of dyons,
i.e.
\begin{align}
\Lambda_{e}\longmapsto & \left[\overrightarrow{\nabla}\cdot\overrightarrow{A}+\frac{1}{c^{2}}\frac{\partial\phi_{e}}{\partial t}\right]=0,\label{eq:35}\\
\Lambda_{m}\longmapsto & \left[\overrightarrow{\nabla}\cdot\overrightarrow{B}+\frac{1}{c^{2}}\frac{\partial\phi_{m}}{\partial t}\right]=0\,.\label{eq:36}
\end{align}
The GDM equations are invariant under the following duality transformations
\begin{align}
\overrightarrow{\mathcal{E}}\longmapsto\,\,c\overrightarrow{\mathcal{H}},\,\,\,\,\,\,\,\,\,\, & \overrightarrow{\mathcal{H}}\longmapsto\,-\overrightarrow{\mathcal{E}}/c,\nonumber \\
\overrightarrow{J}\longmapsto\,\,\overrightarrow{K}/c,\,\,\,\,\,\,\,\,\, & \overrightarrow{K}\longmapsto\,-c\overrightarrow{J},\nonumber \\
\rho_{e}\longmapsto\,\,\rho_{m}/c,\,\,\,\,\,\,\, & \rho_{m}\longmapsto\,-c\rho_{e}.\label{eq:37}
\end{align}
These transformations suggested that the duality transformations of
magnetic and electric current densities follow those of the electric
and magnetic fields, respectively. In GDM equations, the generalized
electric and magnetic field vectors can be expressed in terms of two
four-potentials of dyons,
\begin{alignat}{1}
\overrightarrow{\mathcal{E}} & =-\overrightarrow{\nabla}\phi_{e}-\frac{\partial\overrightarrow{A}}{\partial t}-\overrightarrow{\nabla}\times\overrightarrow{B}\,\,,\label{eq:38}\\
\overrightarrow{\mathcal{H}} & =-\overrightarrow{\nabla}\phi_{m}-\frac{\partial\overrightarrow{B}}{\partial t}+\overrightarrow{\nabla}\times\overrightarrow{A}\,\,.\label{eq:39}
\end{alignat}
Now, the quaternionic vector field $\overrightarrow{\Omega}$ associated
with generalized electromagnetic fields of dyons is defined by
\begin{alignat}{1}
\overrightarrow{\Omega}_{\text{Dyon}} & =\,\,\left(\overrightarrow{\mathcal{E}}+\,i\,\overrightarrow{\mathcal{H}}\right)\;,\,\,i=\sqrt{-1}\,.\label{eq:40}
\end{alignat}
The classical fields equations may also be expressed in terms of two
four-potentials of dyons, i.e.
\begin{align}
\nabla^{2}\overrightarrow{A}-\frac{1}{c^{2}}\frac{\partial^{2}\overrightarrow{A}}{\partial t^{2}}+\overrightarrow{\nabla}\Lambda_{e}+\mu_{0}\overrightarrow{J}= & \,\,0\,,\label{eq:41}\\
\nabla^{2}\overrightarrow{B}-\frac{1}{c^{2}}\frac{\partial^{2}\overrightarrow{B}}{\partial t^{2}}+\overrightarrow{\nabla}\Lambda_{m}-\mu_{0}\overrightarrow{K}= & \,\,0\,,\label{eq:42}\\
\nabla^{2}\phi_{e}-\frac{1}{c^{2}}\frac{\partial^{2}\phi_{e}}{\partial t^{2}}+\frac{\partial\Lambda_{e}}{\partial t}+\frac{\rho_{e}}{\varepsilon_{0}}= & \,\,0\,,\label{eq:43}\\
\nabla^{2}\phi_{m}-\frac{1}{c^{2}}\frac{\partial^{2}\phi_{m}}{\partial t^{2}}+\frac{\partial\Lambda_{m}}{\partial t}+\mu_{0}\rho_{m}= & \,\,0\,.\label{eq:44}
\end{align}
These equations are second order differential wave equations, associated
with two four-current sources to the case of generalized electromagnetic
field of dyons. On the other side, we also may write the electric
and magnetic wave equations for dyons in conducting medium,
\begin{align}
\frac{1}{c^{2}}\frac{\partial^{2}\overrightarrow{\mathcal{E}}}{\partial t^{2}}-\nabla^{2}\overrightarrow{\mathcal{E}}+\mu_{0}\sigma_{e}\frac{\partial\overrightarrow{\mathcal{E}}}{\partial t}+\overrightarrow{\nabla}\left(\frac{\rho_{e}}{\varepsilon_{0}}\right)= & \,\,0\,,\label{eq:45}\\
\frac{1}{c^{2}}\frac{\partial^{2}\overrightarrow{\mathcal{H}}}{\partial t^{2}}-\nabla^{2}\overrightarrow{\mathcal{H}}+\mu_{0}\sigma_{m}\frac{\partial\overrightarrow{\mathcal{H}}}{\partial t}+\overrightarrow{\nabla}\left(\mu_{0}\rho_{m}\right)= & \,\,0\,.\label{eq:46}
\end{align}
where $\sigma_{e}$ and $\sigma_{m}$ are the conductivities due to
electric charge and magnetic monopole, respectively. In generalized
field of dyons, the unified electromagnetic wave equations become
\[
\frac{1}{c^{2}}\frac{\partial^{2}\overrightarrow{\Omega}_{\text{Dyon}}}{\partial t^{2}}-\nabla^{2}\overrightarrow{\Omega}_{\text{Dyon}}+\mu_{0}\sigma\frac{\partial\overrightarrow{\Omega}_{\text{Dyon}}}{\partial t}+\mu_{0}\overrightarrow{\nabla}\varrho=\,\,0\,,
\]
where $\sigma\,(\sigma_{e},\,\sigma_{m})$ is assumed to the conductivity
of dyons and $\varrho\simeq\left(c^{2}\rho_{e}+i\rho_{m}\right)$
is the effective charge density of dyons. We also may write the generalized
vector potential wave equations for dyons in free space, 
\begin{align}
\frac{1}{c^{2}}\frac{\partial^{2}\overrightarrow{A}}{\partial t^{2}}-\nabla^{2}\overrightarrow{A}+\mu_{0}\sigma_{e}\frac{\partial\overrightarrow{A}}{\partial t}+\overrightarrow{\nabla}\Lambda_{e}^{C}= & \,\,0\,,\label{eq:47}\\
\frac{1}{c^{2}}\frac{\partial^{2}\overrightarrow{B}}{\partial t^{2}}-\nabla^{2}\overrightarrow{B}+\mu_{0}\sigma_{m}\frac{\partial\overrightarrow{B}}{\partial t}+\overrightarrow{\nabla}\Lambda_{m}^{C}= & \,\,0\,,\label{eq:48}
\end{align}
and the unified form of wave equations (\ref{eq:47}) and (\ref{eq:48})
become,
\begin{align*}
\frac{1}{c^{2}}\frac{\partial^{2}\overrightarrow{\mathbb{V}}}{\partial t^{2}}-\nabla^{2}\overrightarrow{\mathbb{V}}+\mu_{0}\sigma_{e,m}\frac{\partial\overrightarrow{\mathbb{V}}}{\partial t}+\overrightarrow{\nabla}\Lambda_{e,m}^{C}= & \,\,0\,.
\end{align*}
Here we may identified the Lorentz gauge for conducting medium, respectively
for electric and magnetic charge of dyons, i.e., $\Lambda_{e}^{C}\longmapsto\left(\overrightarrow{\nabla}\cdot\overrightarrow{A}+\frac{1}{c^{2}}\frac{\partial\phi_{e}}{\partial t}+\mu_{0}\sigma_{e}\phi_{e}\right)$
and $\Lambda_{m}^{C}\longmapsto\left(\overrightarrow{\nabla}\cdot\overrightarrow{B}+\frac{1}{c^{2}}\frac{\partial\phi_{m}}{\partial t}+\mu_{0}\sigma_{m}\phi_{m}\right)$.
If we assume the non-conducting media, then the unified vector potential
wave equation for dyons may be expressed as $\square\,\overrightarrow{\mathbb{V}}=0$
with the Lorentz gauge condition $\left(\Lambda_{e}^{C},\,\Lambda_{m}^{C}\right)=\left(\Lambda_{e},\,\Lambda_{m}\right)=0$.

\section{Quantized Maxwell equations for dyons}

We already know that the quantization is a process, which deals the
transition from a classical understanding of physical phenomena to
a newer understanding known as quantum mechanics. In this section,
we have demonstrated the quaternionic generalization of the procedure
for building quantum electrodynamics from classical electrodynamics.
Thus we may compare the classical electromagnetic potential wave equations
to the generalized quaternionic quantum wave equations and obtain
the following equations \cite{key-30,key-31},
\begin{align}
\frac{1}{c^{2}}\frac{\partial^{2}\overrightarrow{A}}{\partial t^{2}}-\nabla^{2}\overrightarrow{A}+2\left(\frac{m_{0}}{\hbar}\right)\frac{\partial\overrightarrow{A}}{\partial t}+\frac{m_{0}^{2}c^{2}}{\hbar^{2}}\overrightarrow{A}= & \,\,0\,.\label{eq:49}\\
\frac{1}{c^{2}}\frac{\partial^{2}\overrightarrow{B}}{\partial t^{2}}-\nabla^{2}\overrightarrow{B}+2\left(\frac{m_{0}}{\hbar}\right)\frac{\partial\overrightarrow{B}}{\partial t}+\frac{m_{0}^{2}c^{2}}{\hbar^{2}}\overrightarrow{B}= & \,\,0\,,\label{eq:50}
\end{align}
These equations are directly governed the quaternionic quantum potential
wave equation for dyons, where we have used the following identities:
\begin{align}
\Lambda_{e}\,\,\, & \longmapsto\,\,\left(\overrightarrow{\nabla}\cdot\overrightarrow{A}+\frac{1}{c^{2}}\frac{\partial\phi_{e}}{\partial t}\right)=-\frac{2m_{0}\phi_{e}}{\hbar}\,,\label{eq:51}\\
\Lambda_{m}\,\,\, & \longmapsto\,\,\left(\overrightarrow{\nabla}\cdot\overrightarrow{B}+\frac{1}{c^{2}}\frac{\partial\phi_{m}}{\partial t}\right)=-\frac{2m_{0}\phi_{m}}{\hbar}\,,\label{eq:52}\\
\overrightarrow{\nabla}\phi_{e} & =-\frac{\partial\overrightarrow{A}}{\partial t}\,,\label{eq:53}\\
\overrightarrow{\nabla}\phi_{m} & =-\frac{\partial\overrightarrow{B}}{\partial t}\,,\label{eq:54}
\end{align}
and the quantized currents for dyons becomes
\begin{align}
\overrightarrow{J} & =\,-\left(\frac{m_{0}^{2}c^{2}}{\mu_{0}\hbar^{2}}\right)\overrightarrow{A}\,.\label{eq:55}\\
\overrightarrow{K} & =\,-\left(\frac{m_{0}^{2}c^{2}}{\mu_{0}\hbar^{2}}\right)\overrightarrow{B}\,.\label{eq:56}
\end{align}
Now, we may interpreted these quantum equations (\ref{eq:51})-(\ref{eq:56})
as fellow:
\begin{itemize}
\item Equations (\ref{eq:51}) and (\ref{eq:52}) are indicated to a novel
equations which show a quantized Lorentz gauge conditions for electric
and magnetic potentials of massive dyons. In other words, we may conclude
that in quantum formulation of generalized electromagnetic fields
of dyons the Lorentz gauge conditions for electric and magnetic potentials
have modified as 
\begin{align*}
\overrightarrow{\nabla}\cdot\overrightarrow{A}+\left(\frac{1}{c^{2}}\frac{\partial}{\partial t}+\frac{2m_{0}}{\hbar}\right)\phi_{e} & =0\,,\\
\overrightarrow{\nabla}\cdot\overrightarrow{B}+\left(\frac{1}{c^{2}}\frac{\partial}{\partial t}+\frac{2m_{0}}{\hbar}\right)\phi_{m} & =0\,.
\end{align*}
\item Equations (\ref{eq:53}) and (\ref{eq:54}) are described to the strength
of longitudinal components of electric and magnetic field vectors,
i.e., $\left(\overrightarrow{\mathcal{E}}_{l}\sim-\overrightarrow{\nabla}\phi_{e}\right)$
and $\left(\overrightarrow{\mathcal{H}}_{l}\sim-\overrightarrow{\nabla}\phi_{m}\right)$.
\item Equations (\ref{eq:55}) and (\ref{eq:56}) are described to the quantized
electric and magnetic current density of dyons. These quantized current
sources are directly depends on the potentials of dyons and also exhibit
the quantum analogue of famous London's equation \cite{key-49} in
presence of electric charge and magnetic monopole.
\end{itemize}
Furthermore, we can express the following quaternionic quantum wave
equations in presence of dual scalar potentials $(\phi_{e},\,\phi_{m})$
for dyons, i.e.,
\begin{align}
\frac{1}{c^{2}}\frac{\partial^{2}\phi_{e}}{\partial t^{2}}-\nabla^{2}\phi_{e}+2\left(\frac{m_{0}}{\hbar}\right)\frac{\partial\phi_{e}}{\partial t}+\frac{m_{0}^{2}c^{2}}{\hbar^{2}}\phi_{e}= & \,\,0\,,\label{eq:57}\\
\frac{1}{c^{2}}\frac{\partial^{2}\phi_{m}}{\partial t^{2}}-\nabla^{2}\phi_{m}+2\left(\frac{m_{0}}{\hbar}\right)\frac{\partial\phi_{m}}{\partial t}+\frac{m_{0}^{2}c^{2}}{\hbar^{2}}\phi_{m}= & \,\,0\:,\label{eq:58}
\end{align}
where
\begin{align}
\rho_{e}= & \,\,-\left(\frac{m_{0}^{2}}{\mu_{0}\hbar^{2}}\right)\phi_{e}\,\,,\label{eq:59}\\
\rho_{m}= & \,\,-\left(\frac{m_{0}^{2}c^{2}}{\mu_{0}\hbar^{2}}\right)\phi_{m}\,\,,\label{eq:60}
\end{align}
are identified the electric and magnetic charge densities of dyons.
Equations (\ref{eq:59}) and (\ref{eq:60}) show a unique quantized
connection between the charge densities to the corresponding scalar
potentials for dyons.\\
In order to write the quantized Proca-Maxwell like equations for massive
dyons in presence of qQM formalism, substituting the quantized transformation
condition given by equation (\ref{eq:17}) along with the quantized
current sources equations of dyons in generalized Dirac-Maxwell's
equation, we establish
\begin{align}
\overrightarrow{\nabla}\cdot\overrightarrow{\mathcal{E}} & -\frac{2m_{0}}{\hbar}\frac{\partial\phi_{e}}{\partial t}-\frac{m_{0}^{2}c^{2}}{\hbar^{2}}\phi_{e}=0\,\,,\label{eq:61}\\
\overrightarrow{\nabla}\cdot\overrightarrow{\mathcal{H}} & -\frac{2m_{0}}{\hbar}\frac{\partial\phi_{m}}{\partial t}-\frac{m_{0}^{2}c^{2}}{\hbar^{2}}\phi_{m}=0\,,\label{eq:62}\\
\overrightarrow{\nabla}\times\overrightarrow{\mathcal{E}} & +\frac{\partial\overrightarrow{\mathcal{H}}}{\partial t}+\frac{m_{0}c^{2}}{\hbar}\overrightarrow{\mathcal{H}}+\frac{2m_{0}}{\hbar}\frac{\partial\overrightarrow{B}}{\partial t}-\frac{m_{0}^{2}c^{2}}{\hbar^{2}}\overrightarrow{B}=0\,,\label{eq:63}\\
\overrightarrow{\nabla}\times\overrightarrow{\mathcal{H}} & -\frac{1}{c^{2}}\frac{\partial\overrightarrow{\mathcal{E}}}{\partial t}-\frac{m_{0}}{\hbar}\overrightarrow{\mathcal{E}}-\frac{2m_{0}}{\hbar}\frac{\partial\overrightarrow{A}}{\partial t}+\frac{m_{0}^{2}c^{2}}{\hbar^{2}}\overrightarrow{A}=0\,.\label{eq:64}
\end{align}
These equations represent a novel approach to the quantized Maxwell
equations for massive particles, called quantized Proca-Maxwell's
(QPM) equations for dyons. However, the Proca equation \cite{key-50}
describes to the classical electrodynamics with finite range (or equivalently
with a non-zero mass). If we introduce a quantum factor, i.e., $\lambda=\frac{h}{m_{0}c}$,
where $\lambda$ is a Compton wavelength for dyons, then the wave
propagation vector ($\overrightarrow{\kappa}$) of dyons become $\overrightarrow{\kappa}=\left(\frac{4\pi^{2}}{\lambda^{2}}\right)^{1/2}\simeq\left(\frac{m_{0}^{2}c^{2}}{\hbar^{2}}\right)^{1/2}$.
Therefore, we may write the compact QPM equations in terms of wave
propagation vector as
\begin{align}
\overrightarrow{\nabla}\cdot\overrightarrow{\mathcal{E}} & =\,\frac{2m_{0}}{\hbar}\frac{\partial\phi_{e}}{\partial t}+\kappa^{2}\phi_{e}\,\,,\label{eq:65}\\
\overrightarrow{\nabla}\cdot\overrightarrow{\mathcal{H}} & =\,\frac{2m_{0}}{\hbar}\frac{\partial\phi_{m}}{\partial t}+\kappa^{2}\phi_{m}\,,\label{eq:66}\\
\overrightarrow{\nabla}\times\overrightarrow{\mathcal{E}}+\frac{\partial\overrightarrow{\mathcal{H}}}{\partial t} & =-\frac{m_{0}c^{2}}{\hbar}\overrightarrow{\mathcal{H}}-\frac{2m_{0}}{\hbar}\frac{\partial\overrightarrow{B}}{\partial t}+\kappa^{2}\overrightarrow{B}\,,\label{eq:67}\\
\overrightarrow{\nabla}\times\overrightarrow{\mathcal{H}}-\frac{1}{c^{2}}\frac{\partial\overrightarrow{\mathcal{E}}}{\partial t} & =\frac{m_{0}}{\hbar}\overrightarrow{\mathcal{E}}+\frac{2m_{0}}{\hbar}\frac{\partial\overrightarrow{A}}{\partial t}-\kappa^{2}\overrightarrow{A}\,.\label{eq:68}
\end{align}
Interestingly, these compact QPM equations of massive dyons are invariant
under the duality transformations, i.e., $\overrightarrow{\mathcal{E}}\longmapsto c\overrightarrow{\mathcal{H}},$
$c\overrightarrow{\mathcal{H}}\longmapsto-\overrightarrow{\mathcal{E}},$
$c\overrightarrow{A}\longmapsto\overrightarrow{B},$ $\overrightarrow{B}\longmapsto-c\overrightarrow{A},$
$\phi_{e}\longmapsto c\phi_{m},$ and $c\phi_{m}\longmapsto-\phi_{e}$.
Thus, the QPM equations are more compact in quaternionic formulation
where the scalar coefficient of quaternionic field visualized to the
scalar electromagnetic field equations and the vector coefficients
of q-algebra visualized to the vector electromagnetic field equations
of massive dyons. Equations (\ref{eq:65})-(\ref{eq:68}) can also
governed to the massless field equations if $m_{0}\sim0$, i.e., 
\begin{align}
\overrightarrow{\nabla}\cdot\overrightarrow{\mathcal{E}} & =\,0\,\,,\label{eq:69}\\
\overrightarrow{\nabla}\cdot\overrightarrow{\mathcal{H}} & =\,0\,,\label{eq:70}\\
\overrightarrow{\nabla}\times\overrightarrow{\mathcal{E}}+\frac{\partial\overrightarrow{\mathcal{H}}}{\partial t} & =0\,,\label{eq:71}\\
\overrightarrow{\nabla}\times\overrightarrow{\mathcal{H}}-\frac{1}{c^{2}}\frac{\partial\overrightarrow{\mathcal{E}}}{\partial t} & =0\,.\label{eq:72}
\end{align}
The above equations are generally described to the macroscopic symmetrical
Maxwell's equations for massless dyons in free space. In the quaternionic
quantum formalism it must be noted that every massive particle will
be contained to a current-source in the quantized electromagnetic
field of dyons. On the other hand, we can say that quantization of
the electromagnetic field of dyons will be occurred only if the particles
have involved some finite mass, otherwise it shows classical field
equations in free space.

\section{Quantized continuity equations and the wave propagation}

In order to establish the quantized continuity equations for the generalized
electromagnetic field of dyons, let us start with the quaternionic
quantized Lorentz gauge condition for electric and magnetic charges
of dyons given by equation (\ref{eq:51}) and (\ref{eq:52}), i.e.,
\begin{align}
\Lambda_{e}^{\mathfrak{Q}}\longmapsto & \,\,\,\,\,\,\overrightarrow{\nabla}\cdot\overrightarrow{A}+\frac{1}{c^{2}}\frac{\partial\phi_{e}}{\partial t}+\frac{2m_{0}\phi_{e}}{\hbar}=\,0\,,\label{eq:73}\\
\Lambda_{m}^{\mathfrak{Q}}\longmapsto & \,\,\,\,\,\,\overrightarrow{\nabla}\cdot\overrightarrow{B}+\frac{1}{c^{2}}\frac{\partial\phi_{m}}{\partial t}+\frac{2m_{0}\phi_{m}}{\hbar}=\,0\,.\label{eq:74}
\end{align}
As such, we have expressed a relation between the classical and quantum
Lorentz gauge conditions of electromagnetic fields for dyons, i.e.,
\begin{align}
\Lambda_{e}^{\mathfrak{Q}}\,= & \,\,\Lambda_{e}+\xi\phi_{e}\,,\label{eq:75}\\
\Lambda_{m}^{\mathfrak{Q}}\,= & \,\,\Lambda_{m}+\xi\phi_{m}\,.\label{eq:76}
\end{align}
where the constant $(\xi\longmapsto2m_{0}/\hbar)$ is a quantum term
depending upon the mass of the particle. As such, for massless dyons,
$\Lambda_{e}^{\mathfrak{Q}}\simeq\Lambda_{e}$ and $\Lambda_{m}^{\mathfrak{Q}}\simeq\Lambda_{m}$.
Therefore, we can summarize that the classical theory shows an approximation
of the quantum theory. By substitute to the quantized value of four-potentials
of dyons, we established the generalized current-source equations
for massive dyons as
\begin{align}
\overrightarrow{\nabla}\cdot\overrightarrow{J}+\frac{\partial\rho_{e}}{\partial t}+\xi\,c^{2}\rho_{e}= & \,\,0\,.\label{eq:77}\\
\overrightarrow{\nabla}\cdot\overrightarrow{K}+\frac{1}{c^{2}}\frac{\partial\rho_{m}}{\partial t}+\xi\,\rho_{m}= & \,\,0\,.\label{eq:78}
\end{align}
These are the quantized continuity equations for massive dyons, where
the additional third terms visualized the torque density result from
the two spin states. Besides, we may obtain the quantized relativistic
wave propagation of massive dyons by the following manner,
\begin{align}
\left(\square-\overrightarrow{\kappa}^{2}\right)\overrightarrow{A}= & \,\,\xi\,\frac{\partial\overrightarrow{A}}{\partial t}\,,\label{eq:79}\\
\left(\square-\overrightarrow{\kappa}^{2}\right)\overrightarrow{B}= & \,\,\xi\,\frac{\partial\overrightarrow{B}}{\partial t}\,,\label{eq:80}\\
\left(\square-\overrightarrow{\kappa}^{2}\right)\phi_{e}= & \,\,\xi\,\frac{\partial\phi_{e}}{\partial t}\,,\label{eq:81}\\
\left(\square-\overrightarrow{\kappa}^{2}\right)\phi_{m}= & \,\,\xi\,\frac{\partial\phi_{m}}{\partial t}\,.\label{eq:82}
\end{align}
Here, equations (\ref{eq:79})-(\ref{eq:82}) are represented the
quantized Klein-Gordon (QKG) field equations for massive dyons. Thus,
the generalized relativistic QKG wave equations are related to the
Schrödinger wave equation in quantum theory. It is second order in
space-time and manifestly invariant under the duality transformations.
Finally,
\begin{align}
\square\overrightarrow{A}= & -\mu_{0}\overrightarrow{J}+\xi\,\frac{\partial\overrightarrow{A}}{\partial t}\,,\label{eq:83}\\
\square\overrightarrow{B}= & -\mu_{0}\overrightarrow{K}+\xi\,\frac{\partial\overrightarrow{B}}{\partial t}\,,\label{eq:84}\\
\square\phi_{e}= & -\frac{\rho_{e}}{\varepsilon_{0}}+\xi\,\frac{\partial\phi_{e}}{\partial t}\,,\label{eq:85}\\
\square\phi_{m}= & -\mu_{0}\rho_{m}+\xi\,\frac{\partial\phi_{e}}{\partial t}\,,\label{eq:86}
\end{align}
are represented the quantized electromagnetic (QEM) potential wave
equations of massive dyons. In four-vectors form, the unified QEM
potential wave equation for dyons become
\begin{align}
\square\mathbb{V}^{\nu}= & -\alpha\mathbb{J}^{\nu}+\xi\,\frac{\partial\mathbb{V}^{\nu}}{\partial t}\,,\label{eq:87}
\end{align}
where $\alpha$ is related to electric and magnetic constants. Here,
we may see that the main distinguish between classical and quantum
formulation is the quantized time varying potential of dyons comes
from the small perturbation terms of time scale given by equation
(\ref{eq:17}). Therefore, we can summarize the quaternionic representation
of classical and the quantum equations for the electromagnetic fields
of dyons given by Table-1.

\begin{table}
\begin{centering}
\textbf{\caption{\textbf{Generalized classical and quantum equations of dyons}}
\medskip{}
}
\par\end{centering}
\centering{}%
\begin{tabular}{ccc}
\hline 
\noalign{\vskip0.3cm}
\textbf{Dyonic fields} & \textbf{Classical equations ($\xi\sim0$)} & \textbf{Quantum equations ($\xi\longmapsto2m_{0}/\hbar$)}\tabularnewline
\hline 
\hline 
\noalign{\vskip0.2cm}
\textbf{$\begin{array}{cc}
\text{Vector}\\
\,\,\,\,\text{waves:}
\end{array}$} & $\begin{cases}
\square\overrightarrow{A}=-\mu_{0}\overrightarrow{J}\\
\square\overrightarrow{B}=-\mu_{0}\overrightarrow{K}
\end{cases}$ & $\begin{cases}
\square\overrightarrow{A}=-\mu_{0}\overrightarrow{J}+\xi\,\frac{\partial\overrightarrow{A}}{\partial t}\,,\\
\square\overrightarrow{B}=-\mu_{0}\overrightarrow{K}+\xi\,\frac{\partial\overrightarrow{B}}{\partial t}\,,
\end{cases}$\tabularnewline
\noalign{\vskip0.2cm}
\textbf{$\begin{array}{cc}
\text{Scalar}\\
\,\,\,\,\text{waves:}
\end{array}$} & $\begin{cases}
\square\phi_{e}=-\frac{\rho_{e}}{\varepsilon_{0}},\\
\square\phi_{m}=-\mu_{0}\rho_{m}
\end{cases}$ & $\begin{cases}
\square\phi_{e}=-\frac{\rho_{e}}{\varepsilon_{0}}+\xi\,\frac{\partial\phi_{e}}{\partial t}\,,\\
\square\phi_{m}=-\mu_{0}\rho_{m}+\xi\,\frac{\partial\phi_{e}}{\partial t}\,,
\end{cases}$\tabularnewline
\noalign{\vskip0.2cm}
\textbf{$\begin{array}{cc}
\text{Four-Potentials}\\
\,\,\,\,\text{waves:}
\end{array}$} & $\square\mathbb{V}^{\nu}=-\alpha\mathbb{J}^{\nu},$ & $\square\mathbb{V}^{\nu}=-\alpha\mathbb{J}^{\nu}+\xi\,\frac{\partial\mathbb{V}^{\nu}}{\partial t}\,,$\tabularnewline
\noalign{\vskip0.2cm}
\textbf{$\begin{array}{cc}
\text{Lorentz}\\
\,\,\text{gauge:}
\end{array}$} & $\begin{cases}
\overrightarrow{\nabla}\cdot\overrightarrow{A}+\frac{1}{c^{2}}\frac{\partial\phi_{e}}{\partial t}=0\\
\overrightarrow{\nabla}\cdot\overrightarrow{B}+\frac{1}{c^{2}}\frac{\partial\phi_{m}}{\partial t}=0
\end{cases}$ & $\begin{cases}
\overrightarrow{\nabla}\cdot\overrightarrow{A}+\frac{1}{c^{2}}\frac{\partial\phi_{e}}{\partial t}+\xi\,\phi_{e}=0\\
\overrightarrow{\nabla}\cdot\overrightarrow{B}+\frac{1}{c^{2}}\frac{\partial\phi_{m}}{\partial t}+\xi\,\phi_{m}=0
\end{cases}$\tabularnewline
\noalign{\vskip0.2cm}
\textbf{$\begin{array}{cc}
\text{Continuity}\\
\text{\,\,equations:}
\end{array}$} & $\begin{cases}
\overrightarrow{\nabla}\cdot\overrightarrow{J}+\frac{\partial\rho_{e}}{\partial t}=0\\
\overrightarrow{\nabla}\cdot\overrightarrow{K}+\frac{1}{c^{2}}\frac{\partial\rho_{m}}{\partial t}=0
\end{cases}$ & $\begin{cases}
\overrightarrow{\nabla}\cdot\overrightarrow{J}+\frac{\partial\rho_{e}}{\partial t}+\xi\,c^{2}\rho_{e}=0\\
\overrightarrow{\nabla}\cdot\overrightarrow{K}+\frac{1}{c^{2}}\frac{\partial\rho_{m}}{\partial t}+\xi\,\rho_{m}=0
\end{cases}$\tabularnewline
\hline 
\end{tabular}
\end{table}

\section{Quantized conductivity in electric and magnetic fields of dyons}

In case of quantized electromagnetic fields of dyons in conducting
media, we may define the microscopic wave propagation of electric
field vector ($\overrightarrow{\mathcal{E}}$) of dyons as
\begin{align}
\frac{1}{c^{2}}\frac{\partial^{2}\overrightarrow{\mathcal{E}}}{\partial t^{2}}-\nabla^{2}\overrightarrow{\mathcal{E}}+2\left(\frac{m_{0}}{\hbar}\right)\frac{\partial\overrightarrow{\mathcal{E}}}{\partial t}+\frac{m_{0}^{2}c^{2}}{\hbar^{2}}\overrightarrow{\mathcal{E}}= & \,\,0\,.\label{eq:88}
\end{align}
Here, equation (\ref{eq:88}) can be obtained by comparing the quaternionic
wave equation (\ref{eq:16}) to the electric wave equation in conducting
medium (\ref{eq:45}). Once, we have obtained the following identities
\begin{align}
2\left(\frac{m_{0}}{\hbar}\right)\,= & \,\,\mu_{0}\sigma_{e}\,,\label{eq:89}\\
\frac{m_{0}^{2}c^{2}}{\hbar^{2}}\overrightarrow{\mathcal{E}}\,= & \,\,\overrightarrow{\nabla}\left(\frac{\rho_{e}}{\varepsilon_{0}}\right)\,.\label{eq:90}
\end{align}
Thus from equations (\ref{eq:89}) and (\ref{eq:90}), we may governed
the following useful relations,
\begin{align}
\sigma_{e}\,= & \,\,\frac{2m_{0}}{\mu_{0}\hbar}\,\simeq\,\,\frac{\xi}{\mu_{0}},\label{eq:91}\\
\overrightarrow{\mathcal{E}}\,= & \,\,\frac{\hbar^{2}}{\varepsilon_{0}m_{0}^{2}c^{2}}\left(\overrightarrow{\nabla}\rho_{e}\right)\,=\,-\overrightarrow{\nabla}\phi_{e}\,\sim\,\overrightarrow{\mathcal{E}}_{l}\,.\label{eq:92}
\end{align}
Equation (\ref{eq:91}) defines to the quantized electrical conductivity
of the electron which is proportional to the mass of the particle,
while equation (\ref{eq:92}) express the effective value of quantized
electric field vector. Similarly, we may write the microscopic wave
equation of magnetic field vector ($\overrightarrow{\mathcal{H}}$)
for dyons as
\begin{align}
\frac{1}{c^{2}}\frac{\partial^{2}\overrightarrow{\mathcal{H}}}{\partial t^{2}}-\nabla^{2}\overrightarrow{\mathcal{H}}+2\left(\frac{m_{0}}{\hbar}\right)\frac{\partial\overrightarrow{\mathcal{H}}}{\partial t}+\frac{m_{0}^{2}c^{2}}{\hbar^{2}}\overrightarrow{\mathcal{H}}= & \,\,0\,,\label{eq:93}
\end{align}
where the quantum identities are governed
\begin{align}
2\left(\frac{m_{0}}{\hbar}\right)\,= & \,\,\mu_{0}\sigma_{m}\,,\label{eq:94}\\
\frac{m_{0}^{2}c^{2}}{\hbar^{2}}\overrightarrow{\mathcal{H}}\,= & \,\,\overrightarrow{\nabla}\left(\mu_{0}\rho_{m}\right)\,.\label{eq:95}
\end{align}
As such, the magnetic conductivity of magnetic monopole and the effective
value of magnetic field vector ($\overrightarrow{\mathcal{H}}$) are
expressed as, respectively,
\begin{align}
\sigma_{m}= & \,\frac{2m_{0}}{\mu_{0}\hbar}\,\,\simeq\,\,\frac{\xi}{\mu_{0}}\,=\,\sigma_{e},\label{eq:96}\\
\overrightarrow{\mathcal{H}}= & -\overrightarrow{\nabla}\phi_{m}\,\sim\,\overrightarrow{\mathcal{H}}_{l}\,.\label{eq:97}
\end{align}
It should be noticed that the value of electric and magnetic conductivity
is same to the case of quantized electric and magnetic fields of dyons.
Moreover, we also may govern the following dynamic equations of massive
dyons for the case of two-four current-sources, i.e.
\begin{equation}
\overrightarrow{\nabla}\rho_{e}+\frac{1}{c^{2}}\frac{\partial\overrightarrow{J}}{\partial t}=\,0\,,\,\,\,\,\overrightarrow{\nabla}\rho_{m}+\frac{\partial\overrightarrow{K}}{\partial t}=\,0\,,\label{eq:98}
\end{equation}
where the dynamic equations of massive dyons have been already discussed
\cite{key-51,key-52}. 

\section{Quantized Dirac wave equations for dyons}

Let us start with the relativistic quantum wave equation \cite{key-53}
or Dirac equation for the particles of mass $m_{0}$ can be expressed
as
\begin{align}
i\hbar\frac{\partial\varPsi}{\partial t}= & \left[c\overrightarrow{\alpha}\cdot\left(-i\hbar\overrightarrow{\nabla}\right)+\beta m_{0}c^{2}\right]\varPsi\,.\label{eq:99}
\end{align}
Here $\overrightarrow{\alpha}$ and $\beta$ are four Hermitian matrices
defined by following $2\times2$ Pauli's $\sigma$-matrices:
\begin{align}
\overrightarrow{\alpha}=\,\begin{pmatrix}0 & \boldsymbol{\sigma}\\
\boldsymbol{\sigma} & 0
\end{pmatrix}\,,\,\,\, & \beta\,=\,\begin{pmatrix}1 & 0\\
0 & -1
\end{pmatrix}\,,\label{eq:100}
\end{align}
where,
\begin{align}
\alpha_{x}^{2}=\alpha_{y}^{2}=\alpha_{z}^{2}=\beta^{2}= & \,\,1\,,\label{eq:101}
\end{align}
and
\begin{align}
\alpha_{x}\alpha_{y}+\alpha_{y}\alpha_{x}= & \,\,\alpha_{y}\alpha_{z}+\alpha_{z}\alpha_{y}=\alpha_{z}\alpha_{x}+\alpha_{x}\alpha_{z}\nonumber \\
= & \,\,\alpha_{x}\beta+\beta\alpha_{x}=\alpha_{y}\beta+\beta\alpha_{y}=\alpha_{z}\beta+\beta\alpha_{z}=0\,.\label{eq:102}
\end{align}
Now, squaring both side of equation (\ref{eq:99}) and obtained
\begin{align}
\left(\overrightarrow{\alpha}\cdot\overrightarrow{\nabla}\right)^{2}\varPsi=\left(-\frac{im_{0}c}{\hbar}\beta-\frac{1}{c}\frac{\partial}{\partial t}\right)^{2}\varPsi & \,\,,\nonumber \\
\frac{1}{c^{2}}\frac{\partial^{2}\varPsi}{\partial t^{2}}-\nabla^{2}\varPsi+\frac{2im_{0}\beta}{\hbar}\frac{\partial\varPsi}{\partial t}-\frac{m_{0}^{2}c^{2}}{\hbar^{2}}\varPsi= & \,\,0\,.\label{eq:103}
\end{align}
Interestingly, equation (\ref{eq:103}) can be written in terms of
quaternionic QEM wave equation for dyons as
\begin{align}
\frac{1}{c^{2}}\frac{\partial^{2}\overrightarrow{\Omega}_{\text{Dyon}}}{\partial t^{2}}-\nabla^{2}\overrightarrow{\Omega}_{\text{Dyon}}+\frac{2M_{eff}}{\hbar}\frac{\partial\overrightarrow{\Omega}_{\text{Dyon}}}{\partial t}+\frac{M_{eff}^{2}c^{2}}{\hbar^{2}}\overrightarrow{\Omega}_{\text{Dyon}}= & \,\,0\,,\label{eq:104}
\end{align}
where we have used
\begin{align}
M_{eff}\,\,\longmapsto & \,\,\left(im_{0}\beta\right).\label{eq:105}
\end{align}
Here the effective mass of dyons ($M_{eff}$) is behave like imaginary
mass (like techyonic dyons). The Bogomolny bound \cite{key-54} shows
the effective mass of dyons which is $M_{eff}\simeq(m_{0}^{e}+i\,m_{0}^{g})$,
where $m_{0}^{e}$ is the mass of electron and $m_{0}^{g}$ is the
mass of magnetic monopole. According to the GUTs, we can estimate
that the mass of monopole \cite{key-55} can be up to an enormous
$10^{17}\,GeV/c^{2}$, which is far greater than the mass of electron
($0.511\,MeV/c^{2}$). Thus we can assumed that the effective mass
of dyons become $M_{eff}\,\simeq\,i\,m_{0}^{g}$.\\
Now, applying four components spinor of state QEM vector $\overrightarrow{\Omega}_{\text{Dyon}}$
associated with two components doublets, i.e., $\overrightarrow{\Omega}_{\text{Dyon}}\longmapsto\begin{pmatrix}\overrightarrow{\Omega}^{+}\\
\overrightarrow{\Omega}^{-}
\end{pmatrix}$. Then the quaternionic quantized Dirac wave equations for dyons can
be written as \cite{key-56,key-57},
\begin{align}
\frac{1}{c^{2}}\frac{\partial^{2}\overrightarrow{\Omega}^{+}}{\partial t^{2}}-\nabla^{2}\overrightarrow{\Omega}^{+}+\frac{2\left(im_{0}\beta\right)}{\hbar}\frac{\partial\overrightarrow{\Omega}^{+}}{\partial t}+\frac{m_{0}^{2}c^{2}}{\hbar^{2}}\overrightarrow{\Omega}^{+}= & \,\,0\,,\label{eq:106}\\
\frac{1}{c^{2}}\frac{\partial^{2}\overrightarrow{\Omega}^{-}}{\partial t^{2}}-\nabla^{2}\overrightarrow{\Omega}^{-}-\frac{2\left(im_{0}\beta\right)}{\hbar}\frac{\partial\overrightarrow{\Omega}^{-}}{\partial t}+\frac{m_{0}^{2}c^{2}}{\hbar^{2}}\overrightarrow{\Omega}^{-}= & \,\,0\,.\label{eq:107}
\end{align}
Thus equations (\ref{eq:106}) and (\ref{eq:107}) are governed two
distinguish energy solutions with ($E\pm m_{0}c^{2}$) and two possible
eigen-states for each of these eigenvalues. Therefore, from these
two novel quantum equations we may concluded that there is the existence
of antiparticle of dyons called \textit{antidyons} \cite{key-58}
(i.e. the composition of antiparticle of electron and antiparticle
of magnetic monopole). We also may analyze the generalized QEM wave
solutions of the Dirac equations for relativistic dyons in free space,
in which an interference phenomena will occurred between positive
and negative energy states of dyons like as the Zitterbewegung wave
equation.

\section{Conclusion }

In this paper, we have extended the formulation of complex algebraic
equations in terms of quaternionic quantum equations for massive dyons.
We have been established a novel approach to qQM, where we have expressed
the various dynamic equations for qQM, viz. the quantum equations
for a moving particles, the unified structure of quaternionic quantum
wave equation and also the quantized transformation condition for
time coordinate. The quaternionic form of quantized potential wave
equations for dyons has been investigated. The qQM formalism described
a novel approach to the Lorentz gauge conditions, respectively, for
the electric charge and magnetic monopoles have also been established.
We have obtained the quantized current sources equations which are
directly depend on the potentials of massive dyons those governed
the quantum analogue of famous London\textquoteright s equation in
presence of dyons. From the qQM formalism, we also have been found
a connection between the quantized electric charge density with the
electric scalar potential and quantized magnetic charge density with
the magnetic scalar potential of dyons. A new form of quantized Proca-Maxwell
equations for dyons has been proposed, which are manifestly invariant
under the duality transformations. Furthermore, we have discussed
that the quaternionic quantization of the electromagnetic field of
dyons will be occurred only if the particles have involved some finite
mass, otherwise it shows the classical field equations in free space.
Accordingly, the quantized continuity equations for generalized electromagnetic
field of massive dyons has been obtained. We have established the
quantized Klein-Gordon like field equations and the unified QEM potential
wave equation for massive dyons. Moreover, the effective quantized
electric and magnetic fields vector has been proposed in a conducting
medium of massive dyons. Finally, we have been investigated the quaternionic
quantized relativistic Dirac wave equations for massive dyons, which
indicated that there will be the existence of antiparticle of dyons
called antidyons. In many theories related to high energy physics
suggested that the dark matter field would be as a source of antiparticles.
Rather, in experimental point of view, it is clear that, like magnetic
monopoles the dyons and antidyons would be too heavy to be produced
at the Large Hadron Collider (LHC). Recently, the experimental study
of magnetic monopole (which created dyons) has been proposed by Daviau
et. al \cite{key-59}. Therefore, we may conclude that, from these
generalized qQM formulation, it will be easy option to establish the
existence of quantum theory for particles viz. dyons, tachyons, axions,
weakly interacting massive particles (WIMPs), etc.

\end{document}